\documentstyle[aps,twocolumn]{revtex}

\begin{document}
\draft

\title{Long beating wavelength in the Schwarz-Hora effect}

\author{Yu. N. Morokov \cite{email}}
\address{
Institute of Computational Technologies, Siberian Branch of \\
the Russian Academy of Sciences, Novosibirsk 630090, Russia}

\maketitle

\begin{abstract}
The quantum-mechanical interpretation of the long-wavelength 
spatial beating of the light intensity in the Schwarz-Hora effect 
is discussed. A more accurate expression for the spatial period 
has been obtained, taking into account the mode structure of the 
laser field within the dielectric film. It is shown that the 
discrepancy of more than 10$\%$ between the experimental and 
theoretical results for the spatial period cannot be reduced by 
using the existing models. More detailed experimental information 
is necessary to clear up the situation. 
\end{abstract}


\narrowtext

In 1969, Schwarz and Hora \cite{1} reported the results of an 
experiment in which a 50-keV beam of electrons passed through a 
thin crystalline film of SiO$_2$, Al$_2$O$_3$, or SrF$_2$ 
irradiated with laser light. Electrons produced the usual 
electron-diffraction pattern at a fluorescent target. However, 
the diffraction pattern was also observed at a nonfluorescent 
target \cite{1,2,3} (the Schwarz-Hora effect). In this case the 
pattern was roughly of the same color as the laser light. The 
effect was absent if the electrical vector of the polarized laser 
light was parallel to the film surfaces. When changing the 
distance between the thin crystalline film and the target, a 
periodic change in the light intensity was observed with spatial 
period of the order of centimeters \cite{2}. The Schwarz-Hora 
effect was discussed extensively in the literature in the early 
1970s. The latest review can be found in Ref. \cite{4}. 

The reported quantitative results \cite{1,2,3,4} were obtained 
for the films of about 1000 $\AA $ thickness. The films were 
illuminated by a 10$^7$-W/cm$^2$ argon ion laser irradiation 
($\lambda _p$= 4880 $\AA $) perpendicular to the electron beam of 
about 0.4 $\mu $A current. These values will be used below for 
numerical estimates.

The quantum-mechanical treatment of the problem was made in the 
one-electron \cite{2,4,5,6,7,8} and many-electron \cite{9,10,11} 
approximations. One problem unresolved up to now is connected 
with the theoretical interpretation of the relatively high 
intensity of the Schwarz-Hora radiation (at least of the order of 
10$^{-10}$ W). The calculated radiated power turns out to be at 
least 10$^3$ times smaller than the observed power 
\cite{4,7,9,10,11,12}. The other problem is connected with the 
strong dependence of the Schwarz-Hora radiation intensity on the 
laser light polarization \cite{2,4,9}. An explanation of this 
dependence is absent too. In the following discussion, we do not 
consider these two problems. 

In this Brief Report we consider only the more transparent 
problem connected with the interpretation of the long-wavelength 
spatial modulation of the Schwarz-Hora radiation 
\cite{2,4,5,6,7,9,10,11,13,14}. The one-particle and 
many-particle models lead to the same expression for the long 
beating wavelength. At first sight, there is even a good 
quantitative agreement with experiment \cite{14}. However, as we 
shall see below, this agreement is accidental. Moreover, there is 
the discrepancy of more than 10$\%$ that cannot be reduced on 
the basis of the existing quantum theories. 

Let the $z$ axis be directed along the incident electron beam. 
The laser beam is along the $x$ axis. The electrical vector of 
the laser light is in the $z$ direction. Electrons pass through 
the dielectric slab restricted by the planes $z = -d$ and $z = 
0$. We consider without loss of generalization only the central 
outgoing electron beam (zeroth-order diffraction). 

Usually the following assumptions are used: An electron interacts 
with the light wave only within the slab; it interacts within the 
slab only with the light wave; the spin effects can be neglected. 
In the simplest case the light field within the slab and 
incident electrons are represented by plane waves. 

Using these assumptions, consider the origin of the 
long-wavelength spatial modulation in the one-electron quantum 
theory. The solution of the Klein-Gordon equation to first order 
in the light field (see, for example, Refs. \cite{5,7}) gives the 
following expression for the electron probability density for $z 
> 0$: 

\begin{eqnarray}
\rho (x,z,t) = \rho _0 \biggl\{ 1-\beta \sin \left[ 
\frac{z}{2\hbar } (2p_0-p_{1z}-p_{-1z})\right] \nonumber \\ 
\times \sin \left( \frac{\pi d}{2d_0}\right) 
\cos \left[ kx - \omega t+\frac{z}{2\hbar } 
(p_{1z}-p_{-1z})\right] \biggl\} .
\label{1}
\end{eqnarray}

Here $\rho _0$ is the probability density for the initial 
incident electron beam and $\omega $ and $k$ denote the circular 
frequency and the wave number of the light wave inside the slab. 
The parameter $\beta $ is proportional to the amplitude of the 
laser field and $d_0$ is the smallest optimum value of the slab 
thickness. For the conditions of the Schwarz experiments, these 
parameters are $\beta $= 0.35 (for $\alpha $-quartz) and $d_0$ = 
1007 $\AA $. The $z$ components of the momentum $p_{nz}$ are 
determined for free electrons of energy $E_n$ and momentum 
${\bbox p}_n$ from the relativistic relationship 

\begin{eqnarray}
E_n^2 = m^2c^4 + {\bbox p}_n^2c^2, 
\label{2}
\end{eqnarray}
\begin{eqnarray*}
E_n = E_0 + n\hbar \omega , \quad 
p_{nx} = n\hbar k, \quad 
n = 0,\pm 1. 
\end{eqnarray*}

Here $m$ is the electron mass. 

The probability that an electron absorbs or emits a photon inside 
the dielectric slab is a periodic function of the slab thickness. 
This is indicated by the second sine term in Eq.\ (\ref{1}). The 
experimental data on such dependence of the Schwarz-Hora 
radiation are absent in the literature. The cosine term 
represents the optical modulation of the electron beam. The 
first sine term in Eq.\ (\ref{1}) is a function of the distance 
$z$ between the slab and the target and represents the stationary 
modulation of the electron probability density. On equating the 
phase of this sine to 2$\pi z/\lambda _b$, we obtain the 
expression for the spatial beating wavelength (the same 
expression is obtained in the many-electron treatment 
\cite{9,10,11}) 

\begin{eqnarray}
\lambda _b = \frac{4\pi\hbar }
{2p_0-p_{1z}-p_{-1z}}. 
\label{3}
\end{eqnarray}

Taking into account Eq.\ (\ref{2}) and that the ratio $\hbar 
\omega /E_0$ is very small, this expression can be rewritten as 
\cite{7} 

\begin{eqnarray}
\lambda _b = \lambda _{b0}\frac{1}{1-(\frac {v_0}{c})^2(1-n^2)}. 
\label{4}
\end{eqnarray}

Here $n = k c/\omega $ is the refractive index of the dielectric 
slab and 

\begin{eqnarray}
\lambda _{b0} = 2\lambda _p\left( \frac{E_0}{\hbar \omega }
\right) \left( \frac{v_0}{c}\right) ^3. 
\label{5}
\end{eqnarray}

It may be assumed that the quantity $E_0-mc^2$ = 50 keV (the 
average energy of incident electrons) was sufficiently well fixed 
in the Schwarz experiments. Therefore, the ratio of the initial 
electron velocity to the velocity of light in vacuum is $v_0/c$ = 
0.4127 and $E_0/\hbar \omega = 2.208\times 10^5$. Then 

\begin{eqnarray}
\lambda _{b0} = 1.515 \enskip {\rm cm}. 
\label{6}
\end{eqnarray}

In the literature, the following three experimental values for 
quantity $\lambda _b$ are presented: 1.70 \cite{2}, 1.75 
\cite{13}, and 1.73$\pm $0.01 \cite{14} cm. 

The authors of Refs. \cite{2,13,14} did not specify for which of 
the three above-mentioned dielectric materials these values had 
been determined. Equation \ (\ref{4}) gives the largest value of 
$\lambda _b$ for strontium fluoride, $\lambda _b=1.29$ cm. This 
material has the smallest value of the refractive index ($n$ = 
1.43) among the three materials used. As affirmed in Ref. 
\cite{4}, the main material used in the experiments was SiO$_2$. 
By using Eq.\ (\ref{4}), we obtain $\lambda _b=1.22$ cm for 
$\alpha $-quartz. Thus it appears that the considered 
quantum-mechanical model does not give the agreement with 
experiment for $\lambda _b$. 

The situation, however, can be somewhat improved. As noted in 
Refs. \cite{6,15}, only one propagation mode of the light wave 
TM$_0$ can be excited within the slab under the experimental 
conditions considered. The corresponding wave field can be 
represented by a superposition of two traveling plane waves, 
propagating at angles $\pm \alpha $ to the $x$ axis. These waves 
turn one into another upon total internal reflection at the slab 
surfaces. The condition for the appearance of the next mode 
TM$_1$ can be written as $d > \lambda _p /2\sqrt{n^2-1}$. For 
$\alpha $-quartz it means $d > 2040\AA $. 

In case the light field is represented by one TM mode, the 
relativistic quantum-mechanical treatment can be carried out by 
analogy with the previous case (see also Ref. \cite{15}). Such 
treatment leads to the same sine term for the stationary spatial 
modulation as that term in Eq.\ (\ref{1}). We obtain the 
following expression for $\lambda _b$: 

\begin{eqnarray}
\lambda _b = \lambda _{b0}\frac{1}{1-(\frac {v_0}{c})^2
(1-n^2\cos ^2\alpha )}. 
\label{7}
\end{eqnarray}

This formula gives a better value for the spatially beating 
wavelength, $\lambda _b = 1.47$ cm, for $\alpha $-quartz if we 
suppose that the light field within the slab is represented by 
the TM$_0$ mode. However, the condition for total internal 
reflection, $n \cos \alpha > 1$, limits the possibility to 
improve the agreement between the theory and experiment by using 
the formula \ (\ref{7}). This implies that $\lambda _b = \lambda 
_{b0}= 1.515$ cm is the upper limit, which cannot be exceeded by 
any formal optimization of the parameters $n$ and $d$. 

Formally, the values $\lambda _b = 1.70-1.75$ cm can be obtained 
by using formula \ (\ref{7}) if we suppose that the dominant role 
in the effect is played by some radiation mode. In this case the 
laser light simply crosses the slab. However, the angles between 
the input laser light and the slab surface must be very large, 
$53-63^{\circ }$, in confrontation with the described 
experimental conditions. 

The wavelength $\lambda _b$ arises in the considered 
quantum-mechanical models as a result of the beat among three 
plane waves representing free electrons. These waves are 
characterized by the quantum numbers $E_n$ and ${\bbox p}_n 
(n=0,\pm 1)$. The values of $E_n$ and $p_{nx}$ are determined 
uniquely by the conservation of energy and the $x$ component of 
quasimomentum in the elementary act of the electron-photon 
interaction inside the dielectric slab. Then the values of 
$p_{nz}$ are determined by the relativistic relationship 
\ (\ref{2}). These factors hold for both the one-particle and 
many-particle considerations. 

Thus the quantity $\lambda _b$ is determined by the simple but 
fundamental propositions of the physical theory. Therefore, we 
can conclude that the quantum models that use the electron 
plane waves ("one-dimensional" in terms of Ref. \cite{16}) have 
no the chance of resolving the discrepancy of more than 10$\%$ 
between theory and experiment for the quantity $\lambda _b$. This 
statement remains valid even if we take into account some 
uncertainty of the published experimental data on the parameters 
$n$ and $d$. 

An attempt to improve the agreement with experiment for 
$\lambda _b$ has been made in Ref. \cite{14}. An expression 
obtained in Ref. \cite{17} was used for a momentum density of a 
light wave in a refracting medium. The agreement has been 
obtained at the cost of repudiating the conservation of the $x$ 
component of quasimomentum in the electron-photon interaction 
inside the slab. However, such a step is incorrect because the 
slab length in the $x$ direction can be considered infinite for 
the conditions of the Schwarz experiments. At the same time, as 
noted in Ref. \cite{17}, the quasimomentum must be conserved in a 
uniform medium. Finally, the formal agreement with experiment 
obtained in Ref. \cite{14} for the case of the plain 
light wave loses any sense for the light field represented by the 
waveguide mode TM$_0$. Calculation shows that the angle $\alpha $ 
is sufficiently large ($\alpha = 46^{\circ }$ for $\alpha 
$-quartz). 

Another contradiction between the theory and experiment can be 
added to the ones noted above. The Schwarz experiments definitely 
indicate \cite{2,16} that there must be the maximum of the 
Schwarz-Hora radiation intensity at the film surface $z$ = 0, 
i.e., there must be the cosine instead of the first sine in 
formula \ (\ref{1}). This problem was discussed in Ref. 
\cite{16}. Then the more rigorous treatment by the same authors 
\cite{11} has in fact confirmed that the theory gives the sine in 
the dependence of the beating effect on the distance $z$. This is 
in accordance also with Ref. \cite{18}. Thus this is one more 
reliably established discrepancy between the theory and the 
experiment. 

In conclusion, the upper limit $\lambda _b$=1.515 cm has been 
obtained for the theoretically permissible values of the 
spatially beating wavelength for the conditions of the Schwarz 
experiments. It does not seem possible to account for the large 
discrepancy between this value and the experimental values 
($\lambda _b^{\rm expt}=1.70-1.75$ cm) on the basis of the 
existing theoretical models. If we add here the other problems 
mentioned above (the radiation intensity, the dependence on laser 
light polarization, and the initial phase of the spatial 
beating), the situation becomes worse. To clear up the situation, 
it is desirable to obtain more detailed experimental information, 
which ought to include, for instance, the dependence of 
$\lambda _b$ on the electron velocity $v_0$ and the refractive 
index of the dielectric film. Unfortunately, the results of the 
Schwarz experiments have not been reproduced by other groups up 
to now. Since 1972 no reports on the results of further attempts 
to repeat those experiments in other groups have appeared while 
the failures of the initial such attempts have been explained by 
Schwarz in Ref. \cite{3}.

\end{document}